\documentclass[prd,showpacs,superscriptaddress,nofootinbib,floatfix,showkeys,10pt]{revtex4}
\usepackage{graphicx}
\usepackage{amsmath}
\usepackage{bm}
\usepackage{yhmath}
\usepackage{mathtools}
\usepackage{wasysym}
\usepackage[colorlinks,citecolor=violet,urlcolor=violet,linkcolor=blue]{hyperref}
\usepackage{subfigure}
\usepackage{color}
\usepackage{cases}
\usepackage{subfigure}
\usepackage{times}
\usepackage{dcolumn,booktabs,bm}
\usepackage{slashed}
\usepackage{amsfonts,amssymb,stmaryrd,latexsym,amsmath}
\usepackage{textcomp}
\usepackage{multirow}
\usepackage{cancel}
\usepackage{array}
\usepackage{orcidlink}
\usepackage{longtable}
\usepackage[utf8]{inputenc}
\usepackage[T1]{fontenc}
\usepackage[utf8]{inputenc}
\usepackage{cprotect}
\usepackage{fvextra}
\usepackage{textcomp}

\begin{document}
\title{Quasitopological Lifshitz dilaton black brane}
\author{A.~Bazrafshan\,}\email{dr.abazrafshan@gmail.com}\thanks{Corresponding Author}
\affiliation{Department of Physics, Jahrom University, Jahrom, P. O. Box 74137-66171, Iran}

\author{M. Ghanaatian\,\orcidlink{0000-0002-7853-6767}}
\affiliation{Department of Physics, Jahrom University, Jahrom, P. O. Box 74137-66171, Iran}

\author{S. Rezaei\,}
\affiliation{Department of  Physics, Payame Noor University (PNU), P. O. Box 19395-3697 Tehran, Iran}

\author{Gh. Forozani\,}
\affiliation{Department of  Physics, Payame Noor University (PNU), P. O. Box 19395-3697 Tehran, Iran}


\begin{abstract}
We construct a new class of $(n+1)$-dimensional Lifshitz dilaton black brane solutions in the presence of the cubic quasitopological gravity for a flat boundary. The related action supports asymptotically Lifshitz solutions by applying some conditions which are used throughout the paper. We have to add a new boundary term and some new counterterms to the bulk action to have finite solutions. Then we define a finite stress tensor complex by which we can calculate the energy density of the quasitopological Lifshitz dilaton black brane. It is not possible to obtain analytical solutions, and so we use some expantions to probe the functions behaviors near the horizon and at the infinity. Combining the equations, we can attain a total constant along the coordinate $r$. At the horizon, this constant is proportional to the product of the temperature and the entropy and at the infinity, the total constant shows the energ density of the quasitopological Lifshitz dilaton black brane. Therefore, we can reach to a relation between the conserved quantities temperature, entropy and the energy density and get a smarr-type formula. Using the first law of thermodynamics, we can find a relation between the entropy and the temperature and then ontain the heat capacity. Our results show that the quasitopological Lifshitz dilaton black brane solutions are thermally stable for each positive values of the dynamical critiacl exponent, $z$.  

\end{abstract}

\pacs{04.70.-s, 04.30.-w, 04.50.-h, 04.20.Jb, 04.70.Bw, 04.70.Dy}

\keywords{Quasitopological gravity; dilaton; Lifshitz. }

\maketitle

\section{Introduction}
At sufficiently high energy scales, gravity is not governed by Einstein action. It is supplied by the string theory which is a promising approach to the quantum gravity and it suggests spacetime dimensions higher than four \cite{Ran,Dvali}. For $n>3$ (where $n$ counts the space dimensions), Einstein equations are not also the most complete ones which satisfy Einstein assumptions. According to Einstein assumption, the left hand side of the field equations is the most general symmetric conserved tensor containing at most second order derivatives of the metric \cite{Kov1,Buch,Lan}. Based on the cosmological point of view, our universe is experiencing a phase of accelerating expansion \cite{Riess,Halver,Tonry,Sperg} which cannot be explained by the standard model of cosmology based on the Einstein theory. Therefore, these reasons have made some physicists to agree that the Einstein theory is not a complete one to describe some of the gravitational and cosmological models.
In order to challenge theses problems, and some others such as the unification of the early-time inflation \cite{Sta} and the late-time cosmic
speed-up \cite{Car,Fay}, one can generalize the Einstein gravity to the  modified gravities. \\
Scalar tensor theory \cite{Amen} is one of the most important modified gravities on which Brans and Dicke theory was the pioneering research. It includes Mach’s principle into gravity in order to study the effects of the scalar fields on the singularity of the universe and to be a candidate for the inflation and quantum gravity \cite{Brans,Brans2}. In the scalar tensor theories, the related action includes one or several long-range scalar fields. Dilaton is one of the scalar fields which is coupled nonminimally to the Einstein gravity at the low-energy limit of the string theory \cite{dilaton1}. In all finite orders of perturbation theory, it is massless \cite{dilaton1}, however a physical dilaton should have mass \cite{Gregory,Gibbons,Horne} in order to avoid the conflict with the classical tests of the gravity tensor character. At the classical level, we can neglect the mass of the dilaton, if the distance scales are small compared to the dilaton compton wavelength.\\ 
Some other kinds of the modified gravities are described by adding some scalars with higher-order curvature corrections to Einstein-Hilbert action. These corrections may lead to the field equations with derivatives of much than 2. Therefore, this is a bonus to construct such modified gravities that result to the field equations with at most second derivatives of metric tensor. \\
Quasitopological gravity \cite{Oliva, Mye} is a modified one with higher-order curvature tensors, where in a spherically symmetry background, it has the
ability to produce second order field equations. At first glance, quasitopological gravity is similar to the LoveLock theory \cite{Love1}, but it has some priorities on the LoveLock theory. For a particular order of the curvature tensor, quasitopological gravity can active gravitationally in smaller dimensions, while the LoveLock theory has no effects on small dimnsions. For example, as the cubic quasitopological gravity is a natural third order curvature
generalization of the Einstein gravity that can cause gravitational effects on five and higher dimensions, third order Lovelock theory is general in just seven and higher dimensions and it does not contribute to the field equations in five dimensions. As this new gravity does not have a topological origin, so it has been called quasitopological gravity \cite{Mye}. Another interesting advantage of the quasitopological gravity originates from causality. In the gauge/gravity duality, the dual CFT should preserve causality. This leads to a constraint on the coupling constants of the gravity theory. In the cubic quasitopological gravity, there are three coupling constants which are determined by the three constraints coming from the positive energy fluxes condition \cite{Buchel,Escobedo,Yang}. So, if one chooses the cubic curvature couplings consistent with these constraints, no evidence for causality violation is observable \cite{Myers2}. For the LoveLock theory, although this accordance is established, but it is not in general, specially for the cases where the gravitational equations of motion are not second order \cite{Hofman}. \\
Quasitopological gravity black hole solutions have been investigated in Refs. \cite{Myers2,Kuang,Brenna1,Brenna,Mehdi,ahmadi,Bazr3}. Lifshitz quasitopological black holes are also in Ref.\cite{Ghan1}. Recently, a study of the Gauss-Bonnet Lifshitz dilaton black brane has been done in Ref. \cite{kord1}. Now, motivated from the mentioned statements, we are eager to consider the cubic quasitopological gravity with the dilaton field in the presence of the Lifshitz spacetime. In this study, due to the higher correction curvature terms that are induced to the Einstein action, one can detect more related quantum gravitational effects. on the other side, study of the quasitopolgical gravity with a scalar field is a new research. Therefore, based on the gauge/gravity duality, these terms can provide a broader backgraound in order to study CFTs with different values for their central charges.\\
This paper is organized as this: \\
In Sec. \eqref{action}, we introduce the bulk action and the related definitions for the cubic quasitopological Lifshitz dilaton black brane in the presence of a massless gauge field. In Sec. \eqref{equation}, we generalize the counterterm
method to the quasitopological gravity and present a finite well-defined action and then vary the related action in order to find the field equations. Using these equations, we can find a total constant along the radial coordinate $r$ in the section \eqref{constant} which can relate us to the thermodynamics. By applying the Lifshitz solution condition in Sec. \eqref{condition}, we can obtain some constraints on the black brane parameters. The study of the black brane behavior at horizon and infinity is in the section \eqref{Ctotal}. We also study the thermodynamics and thermal stability of the quasitopological Lifshitz dilaton black brane in respectively Secs. \eqref{termo} and \eqref{stability}. At last, we have a brief description of the whole paper in Sec. \eqref{result}.
\section{Basic definitions}\label{action}
We start our theory with the $(n+1)$-dimensional cubic quasitopological gravity in the presence of a dilaton field, $\Phi$. So, we consider the following action
\begin{eqnarray}\label{Action}
I_{\mathrm{bulk}}&=&\frac{1}{16\pi}\int_{\mathcal{M}}d^{n+1}x \sqrt{-g}\big(R-2\Lambda+\frac{\lambda}{(n-2)(n-3)}{\mathcal L}_{GB}+\frac{8(2n-1)\mu}{(n-2)(n-5)(3n^2-9n+4)}{\mathcal L}_{TQ}\nonumber\\
&&\,\,\,\,\,\,\,\,\,\,\,\,\,\,\,\,\,\,\,\,\,\,\,\,\,\,\,\,\,\,\,\,\,\,\,\,\,\,\,\,\,\,\,\,\,\,\,\,\,\,\,\,\,\,-\frac{4}{n-1}(\partial_{\mu} \Phi) (\partial^{\mu} \Phi)-e^{-4\alpha \Phi/(n-1)}F_{\mu\nu}F^{\mu\nu}\big),
\end{eqnarray}
where $R$, $\Lambda$ and $\Phi$ show respectively the Einstein-Hilbert Lagrangian, the cosmological constant and the dilaton field. We define a gauge potential $A_{\mu}$, with the definition, $F_{\mu\nu}=\partial_{\mu}A^{\nu}-\partial_{\nu}A^{\mu}$. The constant $\alpha$ can also measure the coupling strength of the dilaton field and the gauge field. The second order LoveLock theory (Gauss-Bonnet) and the cubic quasitopological gravity are respectively showed by ${\mathcal L}_{GB}$ and ${\mathcal L}_{TQ}$ with the definitions
\begin{eqnarray}
&& {\mathcal L}_{GB}= R_{abcd}R^{abcd}-4R_{ab}R^{ab}+R^2\nonumber\\
&& {\mathcal L}_{TQ}=
R_a{{}^c{{}_b{{}^d}}}R_c{{}^e{{}_d{{}^f}}}R_e{{}^a{{}_f{{}^b}}}+\frac{1}{(2n-1)(n-3)}\bigg(\frac{3(3n-5)}{8}R_{abcd}R^{abcd}R-3(n-1)R_{abcd}R^{abc}{{}_e}R^{de}\nonumber\\
&&+3(n+1)R_{abcd}R^{ac}R^{bd}+6(n-1)R_a{{}^b}R_b{{}^c}R_{c}{{}^a}-\frac{3(3n-1)}{2}R_a{{}^b}R_b{{}^a}R +\frac{3(n+1)}{8}R^3\bigg),
\end{eqnarray}
where $\lambda$ and $\mu$ are the related coupling constants.\\ 
We would like to obtain the quasitopological dilaton black brane solutions in the Lifshitz spacetime. Therefore, the $(n+1)$-dimensional metric for the asymptotically Lifshitz solutions with a flat boundary is defined by \cite{Kachru}
\begin{eqnarray}\label{metric1}
ds^2=-\frac{r^{2z}}{l^{2z}}f(r)dt^2+\frac{l^2}{r^2 g(r)}dr^2+r^2 d\vec{X}^2,
\end{eqnarray}
where $z>0$ is a dynamical critical exponent and $l$ defines the AdS radius. $d\vec{X}^2$ represents a $(n-1)$-dimensional hypersurface with a zero curvature boundary and a volume $V_{n-1}$. Lifshitz spacetime has an anisotropic scaling symmetry by
\begin{eqnarray}
t\rightarrow \zeta^{z}t\,\,\, ,\,\,\,r\rightarrow\zeta^{-1}r\,\,\, ,\,\,\,\vec{X}\rightarrow \zeta \vec{X},
\end{eqnarray}
but, it is not conformally invariant. Lifshitz black holes and branes with a dilaton field have been
probed in Ref. \cite{Tarrio}. Also, thermal behavior of charged dilatonic black branes in AdS and UV completions of Lifshitz-like geometries is in Ref. \cite{Bert}. Now we are eager to study the Lifshitz dilaton black brane solutions in the presence of the higher curvature quasitopological gravity. For this purpose, we use the bellow ansatz 
\begin{eqnarray}\label{Amu1}
A_{\mu}=\frac{q}{l^{z}}k(r)\delta_{\mu}^{0}, 
\end{eqnarray}
where $q$ is a constant that will be bounded later .
\section{field equations}\label{equation}
In order to obtain the field equations, we should vary the action \eqref{Action} with respect to the metric tensor, $g_{\mu\nu}$ and the gauge potential, $A_{\mu}$. The variation with respect to $g_{\mu\nu}$ cannot lead to a well-defined variational principle, since we face to a total derivative which has a surface integral with derivative of the metric variation normal to
the boundary, $\partial_{r}\delta g_{\mu\nu}$. In order to eliminate these terms, some surface terms should be added to the action \eqref{Action} which only depend on the geometry of boundary. 
The Gibbons and Hawking surface term is defined by 
\begin{eqnarray}
I^{(1)}_{\mathrm{boundary}}=\frac{1}{8\pi} \int_{\partial \mathcal{M}} d^{n} x \sqrt{-h} K,
\end{eqnarray}
where $h$ is the determinant of the induced metric tensor $h_{\mu\nu}$, on the hypersurface $\partial \mathcal{M}$ at some constant $r$.
$K$ is the trace of the extrinsic curvature, $K_{\mu\nu}=\nabla(_{\mu} n_{\nu})$ where $n_{\nu}$ is the outward-directed unit vector that is orthogonal to the boundary $\partial \mathcal{M}$. The Gauss-Bonnet surface term for the spacetimes with flat boundaries, ($\hat{R}_{abcd}(h)=0$) may be shown by 
\begin{eqnarray}
I^{(2)}_{\mathrm{boundary}}=\frac{1}{8\pi} \int_{\partial \mathcal{M}} d^{n} x \sqrt{-h}\frac{2\lambda}{(n-2)(n-3)} J,
\end{eqnarray}
where $J$ is the trace of the tensor \cite{Deh82,Deh38}
\begin{eqnarray}
J_{ab}=\frac{1}{3}(2KK_{a\gamma}K^{\gamma}_{b}+K_{\gamma\delta}K^{\gamma\delta}K_{ab}-2K_{a\gamma}K^{\gamma\delta}K_{\delta b}-K^2K_{ab}).
\end{eqnarray}
The surface term of the cubic quasitopological gravity for the spacetimes with flat boundaries is also reffered to \cite{Deh44}
\begin{eqnarray}
I^{(3)}_{\mathrm{boundary}}&=&\frac{1}{8\pi} \int_{\partial \mathcal{M}} d^{n} x \sqrt{-h}\,\frac{3\mu }{5n (n-2)(n-1)^2 (n-5)}H,
\end{eqnarray}
where $H$ shows the trace of the tensor $H_{a b}$ that we obtain as bellow tensor  
\begin{eqnarray}\label{Hab}
H_{ab}&=&nK^4 K_{ab}-nK^2 K_{ab}K_{cd}K^{cd}+(n-2)K^3 K_{a}^{c}K_{c b}+(n-1)(3n-2)K_{ac}K^{c}_{b}K_{m d}K^{d}_{e}K^{em}\nonumber\\
&&-3(n-1)(n-2)K_{c d}K^{cd}K_{a e}K^{e m}K_{m b}-n(5n-6)K K_{c d}K^{c d}K_{am}K^{m}_{b}\nonumber\\
&&+(n-1)(5n-6)K K_{a r}K^{c}_{b}K^{r d}K_{c d}.
\end{eqnarray}
So, the total surface term for the action \eqref{Action} is followed by
\begin{eqnarray}
I_{\mathrm{boundary}}=I^{(1)}_{\mathrm{boundary}}+I^{(2)}_{\mathrm{boundary}}+I^{(3)}_{\mathrm{boundary}}.
\end{eqnarray}
Because of the Hamiltonian and other associated thermodynamic quantities, the action $I_{\mathrm{bulk}}+I_{\mathrm{boundary}}$ cannot lead to a finite value when evaluated on the solutions and therefore we should add a local counterterm by \cite{Ross} 
\begin{eqnarray}
I_{\mathrm{ct}}=-\frac{1}{8\pi} \int_{\partial \mathcal{M}} d^{n} x \sqrt{-h} \bigg(\frac{(n-1)(l^4-2\lambda l^2+3\mu)}{l^5}+\frac{2q}{lb^{n-1}}e^{-2\alpha\Phi/(n-1)}(-A_{\gamma}A^{\gamma})^{\frac{1}{2}}\bigg),
\end{eqnarray}
where the constants $q$ and $b$ will be defined later. On the boundary $\partial \mathcal{M}$, the quantity $A_{\alpha}A^{\alpha}=-q^{2}$ is constant for the rotating Lifshitz solutions \cite{Ross}. By these definitions, the total finite action with a well-defined variational principle for the quasitopological dilaton Lifshitz black brane can be manifested by
\begin{eqnarray}\label{Action1}
I_{\mathrm{total}}&=&I_{\mathrm{bulk}}+
I_{\mathrm{boundary}}+I_{\mathrm{ct}}.
\end{eqnarray}
For an ease calculation, we consider the gauge potential \eqref{Amu1} and the metric \eqref{metric1} as bellow
\begin{eqnarray}\label{Amu2}
A_{\mu}=q e^{K(r)} \delta_{\mu}^{0},
\end{eqnarray}
\begin{eqnarray}\label{metric2}
ds^2=-e^{2A(r)}dt^2+e^{2C(r)}dr^2+l^2e^{2B(r)}d\vec{X}^2,
\end{eqnarray}
where we have used the following transformations
\begin{eqnarray}\label{trans1}
A(r)&=&\frac{1}{2} \mathrm{ln}\bigg(\frac{r^{2z}}{l^{2z}}f(r)\bigg),\,\,\,\,\,\,\,\,\,C(r)=-\frac{1}{2} \mathrm{ln}\bigg(\frac{r^{2}}{l^{2}}g(r)\bigg),\nonumber\\
B(r)&=&\mathrm{ln}\bigg(\frac{r}{l}\bigg),\,\,\,\,\,\,\,\,\,\,\,\,\,\,\,\,\,\,\,\,\,\,\,\,\,\,\,\,\,K(r)= \mathrm{ln}\bigg(\frac{k(r)}{l^z}\bigg).
\end{eqnarray}
If we substitute the relations \eqref{Amu2} and \eqref{metric2} in the action \eqref{Action} and then integrate by part, we get to a Lagrangian where by varying it with respect to the functions $A(r)$, $B(r)$, $C(r)$, $\Phi(r)$ and $K(r)$, the field equations are obtained as bellow
\begin{eqnarray}\label{E1}
E_{1}&=& [B^{''}+\frac{n}{2}B^{'2}-B^{'}C^{'}+\frac{2}{(n-1)^2}\Phi^{'2}]e^{A-C+(n-1)B}-\frac{n}{2}\lambda [\frac{4}{n}B^{''}B^{'2}+B^{'4}-\frac{4}{n}B^{'3}C^{'}]e^{A-3C+(n-1)B}\nonumber\\
&&-\frac{n}{2}\mu [-\frac{6}{n}B^{'4}B^{''}+\frac{6}{n}B^{'5}C^{'}-B^{'6}]e^{A+(n-1)B-5C}
+\frac{\Lambda}{n-1}e^{A+C+(n-1)B}\nonumber\\
&&+\frac{q^2}{n-1}K^{'2}e^{-A-C+(n-1)B+2K-4\alpha\Phi/(n-1)}=0,
\end{eqnarray}
\begin{eqnarray}\label{E2}
E_{2}&=&-[-(n-2)B^{''}-A^{''}-\frac{(n-1)(n-2)}{2}B^{'2}-A^{'2}+(n-2)B^{'}(C^{'}-A^{'})+A^{'}C^{'}-\frac{2}{n-1}\Phi^{'2}]e^{A-C+(n-1)B}\nonumber\\
&&-2\lambda B^{'}
\bigg\{2A^{'}B^{''}+B^{'}\bigg[B^{'}\bigg(\frac{(n-1)(n-4)}{4}B^{'}+(n-2)A^{'}-(n-4)C^{'}\bigg)+A^{'}(A^{'}-3C^{'})+(n-4)B^{''}+A^{''}\bigg]\bigg\}  \nonumber\\
&&\times e^{A-3C+(n-1)B}+e^{A+C+(n-1)B}\Lambda-q^2 K^{'2}e^{-A-C+(n-1)B+2K-4\alpha\Phi/(n-1)}
\nonumber\\
&&+3\mu B^{'3}\bigg\{4A^{'}B^{''}+B^{'}\bigg[B^{'}\bigg(\frac{(n-1)(n-6)}{6}B^{'}+(n-2)A^{'}-(n-6)C^{'}\bigg)+A^{'}(A^{'}-5C^{'})+(n-6)B^{''}+A^{''}\bigg]\bigg\}\nonumber\\
&&\times e^{A+(n-1)B-5C}=0,
\end{eqnarray}
\begin{eqnarray}\label{E3}
E_{3}&=& [A^{'}B^{'}+\frac{n-2}{2}B^{'2}-\frac{2}{(n-1)^2}\Phi^{'2}]e^{A-C+(n-1)B}-2\lambda B^{'3} [A^{'}+\frac{n-4}{4}B^{'}] e^{A-3C+(n-1)B}+\frac{\Lambda}{n-1}e^{A+C+(n-1)B}\nonumber\\
&&+3\mu B^{'5} [A^{'}+\frac{n-6}{6}B^{'}]e^{A+(n-1)B-5C}+\frac{q^2}{n-1} K^{'2}e^{-A-C+(n-1)B+2K-4\alpha\Phi/(n-1)}=0,
\end{eqnarray}
\begin{eqnarray}\label{E4}
E_{4}&=& -\frac{4}{(n-1)^2}\{[\Phi^{''}+\Phi^{'}(A^{'}-C^{'}+(n-1)B^{'})]e^{A-C+(n-1)B}-\alpha q^2 K^{'2}e^{-A-C+(n-1)B+2K-4\alpha\Phi/(n-1)}\}=0,
\end{eqnarray}
\begin{eqnarray}\label{E5}
E_{5}&=& \frac{2q^2}{(n-1)}\bigg\{K^{''}+K^{'2}+K^{'}\bigg[-\frac{4}{n-1}\alpha\Phi^{'}-A^{'}-C^{'}+(n-1)B^{'}\bigg]\bigg\}e^{-A-C+(n-1)B+2K-4\alpha\Phi/(n-1)}=0.
\end{eqnarray}
In the above equations, the prime $(')$ is the derivative with respect to the $r$ coordinate. We also mention that by choosing $\mu=0$, we get to the results of the Gauss-Bonnet-dilaton Lifshitz black brane in Ref.\cite{kord1}. Under some conditions, we can get to the exact solutions. In the absence of the higher curvature tensors ($\mu=\lambda=0$), we obtain the exact Lifshitz-dilaton black brane solutions in Ref. \cite{Tarrio}. For $z=1$ and in the absence of the dilaton field, $\Phi(r)=0$, the solutions are the ones in Ref. \cite{ahmadi}. \\
As we have gotten to some coupled differential equations in Eqs.\eqref{E1}-\eqref{E5}, so it is not possible to gain an analytic solution for a general case, $\mu$, $\lambda$, $\Phi\neq0$ and $z\neq 1$. In the next section, we obtain some constants that can lead us to the conserved quantities. \\
\section{constants along the radial coordinate }\label{constant}
Now we want to gain the constants which may lead us to the conserved quantities. If we combine the equations, we get to
\begin{eqnarray}
E_{1}-\frac{E_{2}}{n-1}+E_{5}&=&-\frac{1}{n-1}\bigg[e^{A-C+(n-1)B}(A^{'}-B^{'})+2\lambda e^{A-3C+(n-1)B}(B^{'3}-A^{'}B^{'2})-3\mu e^{A+(n-1)B-5C}(B^{'5}-A^{'}B^{'4})\nonumber\\
&&-2q^2 e^{-A-C+(n-1)B+2K-4\alpha\Phi/(n-1)}K^{'}\bigg]^{'}=0,
\end{eqnarray}
and
\begin{eqnarray}
E_{4}+\frac{2}{n-1}\alpha E_{5}=-\frac{4}{(n-1)^2}\bigg[e^{A-C+(n-1)B}\Phi^{'}-\alpha q^2 e^{-A-C+(n-1)B+2K-4\alpha \Phi/(n-1)} K^{'}\bigg]^{'}=0,
\end{eqnarray} 
where they result to two constants $C_{1}$ and $C_{2}$ along the radial coordinate r as bellow 
\begin{eqnarray}\label{C10}
C_{1}&=&-\frac{1}{n-1}[e^{A-C+(n-1)B}(A^{'}-B^{'})+2\lambda e^{A-3C+(n-1)B}(B^{'3}-A^{'}B^{'2})-3\mu e^{A+(n-1)B-5C}(B^{'5}-A^{'}B^{'4})\nonumber\\
&&-2q^2 e^{-A-C+(n-1)B+2K-4\alpha\Phi/(n-1)}K^{'}],
\end{eqnarray}
\begin{eqnarray}\label{C20}
C_{2}=-\frac{4}{(n-1)^2}\bigg[e^{A-C+(n-1)B}\Phi^{'}-\alpha q^2 e^{-A-C+(n-1)B+2K-4\alpha \Phi/(n-1)} K^{'}\bigg].
\end{eqnarray} 
By replacing the transformations \eqref{trans1} in Eq. \eqref{E5} and then solving this equation, we can find a relation between the theory functions by
\begin{eqnarray}\label{maxwell2}
\frac{d}{dr} k(r)=r^{z-n}\sqrt{\frac{f(r)}{g(r)}} e^{4\alpha\Phi(r)/(n-1)}.
\end{eqnarray}
So, if we use the transformations \eqref{trans1} and Eq.\eqref{maxwell2} in the constants \eqref{C10} and \eqref{C20}, we can obtain these constants as  
\begin{eqnarray}
C_{1}&=&-\frac{1}{2(n-1) l^{z+n}}\bigg\{\bigg[1-\frac{2\lambda}{l^2}g+\frac{3\mu}{l^4}g^2\bigg]\bigg[r^{z+n}f^{'}\sqrt{\frac{g}{f}}+2r^{z+n-1}(z-1)\sqrt{fg}\bigg]-4q^2k\bigg\},
\end{eqnarray}
and
\begin{eqnarray}
C_{2}=-\frac{4}{(n-1)^2 l^{z+n}}\bigg[r^{z+n}\sqrt{fg}\Phi^{'}-\alpha q^2 k\bigg].
\end{eqnarray}
We can also combine the constants $C_{1}$ and $C_{2}$ to get a new total constant 
\begin{eqnarray}\label{totalC}
C_{\mathrm{total}}=-2(n-1)l^{n-1}\bigg(C_{1}-\frac{n-1}{2\alpha}C_{2}\bigg)=\frac{r^{z+n}}{l^{z+1}} \bigg\{\bigg(1-\frac{2\lambda}{l^2}g+\frac{3\mu}{l^4}g^2\bigg)\bigg[f^{'}\sqrt{\frac{g}{f}}+\frac{2(z-1)}{r}\sqrt{fg}\bigg]-\frac{4\Phi^{'}}{\alpha}\sqrt{fg}\bigg\},
\end{eqnarray}
where it will be used in Sec.\eqref{termo}. For $z=1$ (where $f(r)=g(r)$) and in the absence of the dilaton field, $\Phi(r)=0$, the total constant reduces to  
\begin{eqnarray}\label{totalC1}
C_{\mathrm{total}}=\frac{r^{n+1}}{l^{2}} \bigg\{\bigg(f-\frac{\lambda}{l^2}f^2+\frac{\mu}{l^4}f^3\bigg)^{'}\bigg\},
\end{eqnarray}
where it is proportional to the mass of the quasitopological black brane with $s=1$ in Ref. \cite{ahmadi}.
\section{Conditions for Lifshitz solutions}\label{condition}
Now, it's the time to check the possibility of the dilaton Lifshitz black brane solutions in the presence of the cubic quasitopological gravity. This leads to get to some constraints on the theory parameters where they are the prerequisites for the Lifshitz quasitopological dilaton black brane solutions. The action \eqref{Action} supports solutions asymptotic to the Lifshitz solutions, if we consider $f(r)=g(r)=1$ where
\begin{eqnarray}\label{lifmet}
ds^2=-\frac{r^{2z}}{l^{2z}}dt^2+\frac{l^2}{r^2}dr^2+r^2d\vec{X}^2.
\end{eqnarray}
Using the metric \eqref{lifmet}, we probe the possibility of the Lifshitz quasitopological dilaton black brane solutions with and without matter sources. 
\subsection{Matter-free solutions}
If we apply the metric \eqref{lifmet} and the matter-free condition ($k(r)=\Phi(r)=0$) in Eqs. \eqref{E1}-\eqref{E4}, they reduce to the bellow conditions
\begin{eqnarray}\label{con1}
\lambda=2l^2\bigg[1+\frac{3\Lambda}{n(n-1)}l^2\bigg]\,\,,\,\,\mu=l^4\bigg[1+\frac{4\Lambda}{n(n-1)}l^2\bigg],
\end{eqnarray}
where they are independent of the critical exponent $z$. In the absence of the matter field, the conditions in \eqref{con1} are the possibility of the quasitopological Lifshitz solutions which should be implied on the related equations. Imposing the conditions \eqref{con1} for the case $f(r)=g(r)$ in Eq. \eqref{E1}, we get to 
\begin{eqnarray}
\frac{\Lambda l^6}{n}(1-3g(r)^2+2g(r)^3)+\frac{(n-1)l^4}{2}(g(r)-2g(r)^2+g(r)^3)=\frac{C}{r^n},
\end{eqnarray}
where $C$ is a constant of integration. This third-order equation may have exact solutions. It should be also mentioned that by choosing $C=0$, we have the exact solution $g(r)=1$. 
\subsection{Matter solutions}
Now, we aim to obtain the conditions for the existence of the asymptotically Lifshitz solutions in quasitopological gravity with matter fields, $k(r)\neq1$ and $\Phi(r)\neq0$. So, if we use the relation $f(r)=g(r)=1$ in Eqs. \eqref{E1}-\eqref{E4}, they lead to
\begin{eqnarray}\label{E11}
-r^2 \Phi^{'2}-\frac{n(n-1)^2}{4}\bigg(1-\frac{\lambda}{l^2}+\frac{\mu}{l^4}\bigg)-\frac{n-1}{2}\bigg[\Lambda l^2+\frac{q^2 e^{4\alpha\Phi/(n-1)}}{r^{2(n-1)}}\bigg]=0,
\end{eqnarray}
\begin{eqnarray}\label{E22}
r^2 \Phi^{'2}+\frac{(n-1)}{4}\bigg[(n+z-1)(n+z-2)+z(z-1)-\frac{\lambda}{l^2}((n+z-1)(n+3z-4)+z(z-1))\nonumber\\+\frac{\mu}{l^4}\bigg((n+z-1)(n+5z-6)+z(z-1)\bigg)\bigg]+\frac{n-1}{2}\bigg[\Lambda l^2-\frac{q^2 e^{4\alpha\Phi/(n-1)}}{r^{2(n-1)}}\bigg]=0,
\end{eqnarray}
\begin{eqnarray}\label{E33}
r^2 \Phi^{'2}-\frac{(n-1)^2}{4}\bigg[n+2(z-1)-\frac{\lambda}{l^2}\bigg(n+4(z-1)\bigg)+\frac{\mu}{l^4}\bigg(n+6(z-1)\bigg)\bigg]-\frac{n-1}{2}\bigg[\Lambda l^2+\frac{q^2 e^{4\alpha\Phi/(n-1)}}{r^{2(n-1)}}\bigg]=0,
\end{eqnarray}
\begin{eqnarray}\label{E44}
r \Phi^{''}+(n+z)\Phi^{'}-\frac{\alpha q^2 e^{4\alpha\Phi/(n-1)}}{r^{2n-1}}=0.
\end{eqnarray}
By subtracting Eq. \eqref{E33} from Eq. \eqref{E11}, we can solve the obtained equation to get to  
\begin{eqnarray}\label{philif}
\Phi(r)=\frac{n-1}{2}\sqrt{(z-1)(1-\frac{2\lambda}{l^2}+\frac{3\mu}{l^4})}\,\, \mathrm{ln}(\frac{r}{b}),
\end{eqnarray}
where $b$ is a constant of integration. Now if we use the condition \eqref{philif} in Eqs. \eqref{E11}-\eqref{E44}, we may reach some constraints on the parameters $\Lambda$, $q$, and $\alpha$ as bellow 
\begin{eqnarray}\label{lambda1}
\Lambda=-\frac{1}{2l^{2}}[z^2+(2n-3)z]\bigg[1-\frac{2\lambda}{l^2}+\frac{3\mu}{l^4}\bigg]-\frac{n-1}{2l^2}\bigg[n-2-\frac{n-4}{l^2}\lambda+\frac{n-6}{l^4}\mu\bigg],
\end{eqnarray}
\begin{eqnarray}\label{q1}
q=\frac{b^{n-1}}{\sqrt{2}}\sqrt{(z-1)(z+n-1)(1-\frac{2\lambda}{l^2}+\frac{3\mu}{l^4})},
\end{eqnarray}
\begin{eqnarray}\label{alpha1}
\alpha=\frac{n-1}{\sqrt{(z-1)(1-\frac{2\lambda}{l^2}+\frac{3\mu}{l^4})}}.
\end{eqnarray}
The constrained value of the parameter $q$ in Eq. \eqref{q1} shows that $q$ is not a free parameter and we cannot consider it an electric charge. So we have an uncharged quasitopological Lifshitz dilaton black brane. 
It is also notable that for $z=1$, the parameter $q$ and the dilaton field $\Phi$ vanish and so an uncharged quasitopological black brane is emerged \cite{Mye}. In the absence of the cubic quasitopological gravity with $\mu=0$, the constraints reduce to the ones in the gauss-bonnet-dilaton Lifshitz black brane \cite{kord1}. 
It is also clear from the conditions \eqref{q1} and \eqref{alpha1} that the bellow condition should be satisfied
\begin{eqnarray}
1-\frac{2\lambda}{l^2}+\frac{3\mu}{l^4}>0. 
\end{eqnarray}
Now if $\lambda>0$, we can deduce that
\begin{eqnarray}\label{aaa}
-\frac{1}{2l^6}\big\{\big(l^4+3\mu\big)\big[z(z-1)+2z(n-1)+(n-1)(n-2)\big]-2n(n-1)\mu\big\}\leq \Lambda<-\frac{n(n-1)}{4l^6}\big(l^4-\mu\big).
\end{eqnarray}
It is clear from the relation \eqref{aaa} that for $\mu<l^4$, the cosmological constant is negative ($\Lambda<0$) and for $\mu>l^4$, it is positive if  
\begin{eqnarray}
\frac{\big(l^4+3\mu\big)\big[z(z-1)+2z(n-1)+(n-1)(n-2)\big]-2n(n-1)\mu}{[n(n-1)+2(z-1)(z+2n-2)]l^4}<\frac{\lambda}{l^2}<\frac{1}{2}+\frac{3\mu}{2l^4}.
\end{eqnarray}
\section{asymptotic behaviors}\label{Ctotal}
We concluded that it is impossible to obtain exact Lifshitz solutions of the quasitopological dilaton black brane. So, it can be useful to study the solutions behaviors near the horizon and at the infinity. 
\subsection{near the horizon}\label{horizon}
We know that for a nonextreme black brane,
$f(r_{+})$ and $g(r_{+})$ go to zero linearly, where $r_{+}$ is the event horizon radius. So, we expand the functions $f(r)$, $g(r)$ and $\Phi(r)$ near $r=r_{+}$ as bellow
\begin{eqnarray}\label{basthorizon}
f(r)&=& f_{1}\{(r-r_{+})+f_{2} (r-r_{+})^2+f_{3} (r-r_{+})^3+...\}\nonumber\\
g(r)&=& g_{1}(r-r_{+})+g_{2} (r-r_{+})^2+g_{3} (r-r_{+})^3+...\nonumber\\
\Phi(r)&=& \Phi_{+}+\Phi_{1}(r-r_{+})+\Phi_{2} (r-r_{+})^2+\Phi_{3} (r-r_{+})^3... .
\end{eqnarray}
where the coefficients $f_{i}$'s, $g_{i}$'s and $\Phi_{i}$'s are constant. It is also remarkable that the dilaton field has a nonzero value at the horizon. If we replace the above expansions into the equations \eqref{E1}-\eqref{E5} and also use the constraints \eqref{philif}-\eqref{alpha1}, we can solve the equations for each powers of $(r-r_{+})$ and then obtain the all coefficients in terms
of $r_{+}$.\\
If we substitute the expansions \eqref{basthorizon} in the relation \eqref{totalC}, we can get to the total constant at the horizon as bellow
\begin{eqnarray}\label{Cth}
C_{\mathrm{total}}=\frac{r_{+}^{z+n}}{l^{z+1}}\sqrt{f_{1} g_{1}}.
\end{eqnarray}
\subsection{at the infinity}\label{infinity}
In this part, we aim to probe the behavior of the quasitopological Lifshitz dilaton black brane at the infinity. For this purpose, we use the straightforward perturbation theory by 
\begin{eqnarray}\label{exfinity}
f(r)&=& 1+\epsilon f_{1}(r),\nonumber\\
g(r)&=& 1+\epsilon g_{1}(r),\nonumber\\
\Phi(r)&=& \Phi_{0}(r)+\epsilon \Phi_{1}(r),
\end{eqnarray}
where $\Phi_{0}(r)$ has defined in Eq.\eqref{philif}.  
If we replace the expansions \eqref{exfinity} and the conditions \eqref{philif}-\eqref{alpha1} in the equations \eqref{E1} -\eqref{E4}, then we can find them up to the first order in $\epsilon$ as bellow 
\begin{eqnarray}\label{Eas1}
&&(n-1)(l^4-2\lambda l^2+3\mu)r g_{1}^{'}+4l^2\sqrt{(z-1)(l^4-2\lambda l^2+3\mu)}\,[r\Phi_{1}^{'}+(n-1+z)\Phi_{1}]+\nonumber\\
&&(n-1)(n+z-1)(l^4-2\lambda l^2+3\mu)g_{1}=0,
\end{eqnarray}
\begin{eqnarray}
&&(l^4-2\lambda l^2+3\mu)r^2f_{1}^{''}+(n+2z-1)(l^4-2\lambda l^2+3\mu)rf_{1}^{'}+4l^2 \sqrt{(z-1)(l^4-2\lambda l^2+3\mu)}\,r\Phi_{1}^{'}+\nonumber\\
&&[(n+z-2)l^4-2\lambda(n+3z-4)l^2+3\mu(n+5z-6)]\,rg_{1}^{'}+(n+z-1)[(n+2z-3)l^4-2\lambda(n+4z-5)l^2\nonumber\\
&&+3\mu(n+6z-7)]g_{1}-4l^2(n+z-1)\sqrt{(z-1)(l^4-2\lambda l^2+3\mu)}\,\Phi_{1}=0
\end{eqnarray}
\begin{eqnarray}
&&(n-1)(l^4-2\lambda l^2+3\mu)rf_{1}^{'}-4l^2\sqrt{(z-1)(l^4-2\lambda l^2+3\mu)}\,r\Phi_{1}^{'}+4l^2(n+z-1)\sqrt{(z-1)(l^4-2\lambda l^2+3\mu)}\,\Phi_{1}\nonumber\\
&&+(n-1)[(n+z-1)l^4-2\lambda l^2(n+3z-3)+3\mu(n+5z-5)]g_{1}=0,
\end{eqnarray}
\begin{eqnarray}\label{Eas4}
&&4l^2 r[r\Phi_{1}^{''}+(n+z)\Phi_{1}^{'}]+(n-1)\sqrt{(z-1)(l^4-2\lambda l^2+3\mu)}[r(f_{1}^{'}+g_{1}^{'})+2(n+z-1)g_{1}]\nonumber\\&&-8l^2(n-1)(n+z-1)\Phi_{1}=0.
\end{eqnarray}
It is a matter of calculation to show that the suitable solutions for the equations \eqref{Eas1}-\eqref{Eas4} are chosen by 
\begin{eqnarray}\label{soluinf}
f(r)&=& 1+\epsilon\bigg\{-\frac{C_{1}}{r^{n+z-1}}-\frac{C_{2}}{r^{(n+z+\gamma-1)/2}}-\frac{C_{3}}{r^{(n+z-\gamma-1)/2}}\bigg\},\nonumber\\
g(r)&=& 1+\epsilon\bigg\{-\frac{a_{1} C_{1}}{r^{n+z-1}}-\frac{a_{2} C_{2}}{r^{(n+z+\gamma-1)/2}}-\frac{a_{3}C_{3}}{r^{(n+z-\gamma-1)/2}}\bigg\},\nonumber\\
\Phi(r)&=&\frac{n-1}{2}\sqrt{(z-1)(1-\frac{2\lambda}{l^2}+\frac{3\mu}{l^4})}\,\, \mathrm{ln}\bigg(\frac{r}{b}\bigg)+\epsilon\bigg\{-\frac{b_{1} C_{1}}{r^{n+z-1}}-\frac{b_{2} C_{2}}{r^{(n+z+\gamma-1)/2}}-\frac{b_{3}C_{3}}{r^{(n+z-\gamma-1)/2}}\bigg\},
\end{eqnarray}
where the related parameters are
\begin{eqnarray}
a_{1}&=&(n+z-1)(n+z-2)(l^4-2\lambda l^2+3\mu)\mathcal{K}^{-1},\nonumber\\
a_{2}&=&-(n+z+\gamma-1)(l^4-2\lambda l^2+3\mu)\mathcal{K}_{+}^{-1},\nonumber\\
a_{3}&=&-(n+z-\gamma-1)(l^4-2\lambda l^2+3\mu)\mathcal{K}_{-}^{-1},\nonumber\\
b_{1}&=&\frac{n-1}{2l^2}(z-1)(\lambda l^2-3\mu)\sqrt{(z-1)(l^4-2\lambda l^2+3\mu)}\,\mathcal{N}_{+}^{-1},\nonumber\\
b_{2}&=&\frac{n-1}{4 l^2}(n+z+\gamma-1)(l^4-2\lambda l^2+3\mu)\sqrt{\frac{(l^4-2\lambda l^2+3\mu)}{z-1}}\mathcal{K}_{+}^{-1},\nonumber\\
b_{3}&=&\frac{n-1}{4 l^2}(n+z-\gamma-1)(l^4-2\lambda l^2+3\mu)\sqrt{\frac{(l^4-2\lambda l^2+3\mu)}{z-1}}\mathcal{K}_{-}^{-1},\nonumber\\
\mathcal{K}&=&(n+z-1)(n+z-2)l^4-2\lambda l^2[(n+z-1)(n+z-4)+2nz]+3\mu(n+z-1)(n+z-6)+4nz,\nonumber\\
\mathcal{K}_{\pm}&=&(n+z\pm\gamma-1)l^4-2\lambda l^2(n\pm\gamma-3z+3)+3\mu(n\pm\gamma-7z+7),\nonumber\\
\mathcal{N}_{+}&=&(n+z-1)(n+z-2)l^4-2\lambda l^2[n^2+(4z-5)n+(z-1)(z-4)]+3\mu[n^2+(6z-7)n+(z-1)(z-6)],\nonumber\\
\gamma^2&=&[(n+z-1)(9n+9z-17)l^4-2\lambda l^2(9n^2+2(9z-13)n+(z-1)(z-9))\nonumber\\
&&+3\mu(9n^2+2(9z-13)n-(7z+1)(z-1))][l^4-2\lambda l^2+3\mu]^{-1}.\nonumber\\
\end{eqnarray}
At the limit $r\rightarrow\infty$, the functions $f(r)$ and $g(r)\rightarrow 1$ and $\Phi(r)$ goes to the value in Eq.\eqref{philif}. In order to have this asymptotically behavior for the above functions, we should choose $C_{3}=0$ \cite{Deh38}. 
Now, if we use the above obtained solutions in the relation \eqref{totalC}, the total contant at the infinity is earned as bellow 
\begin{eqnarray}\label{Ctinf}
C{\mathrm{total}}=\frac{(n+z-1)(l^4-2\lambda l^2+3\mu)[(n+z-2)(n+z-1)(l^4-2\lambda l^2+3\mu)+2(z-1)^2(\lambda l^2-3\mu)]}{\mathcal{N}_{+}l^{z+5}}C_{1}.
\end{eqnarray}
We can see that at the infinity, only the constant $C_{1}$ is effective on the constant $C_{\mathrm{total}}$ and $C_{2}$ has no effect.\\
\section{Thermodynamics and conserved quantities}\label{termo}
\begin{figure}
\center
\includegraphics[scale=0.5]{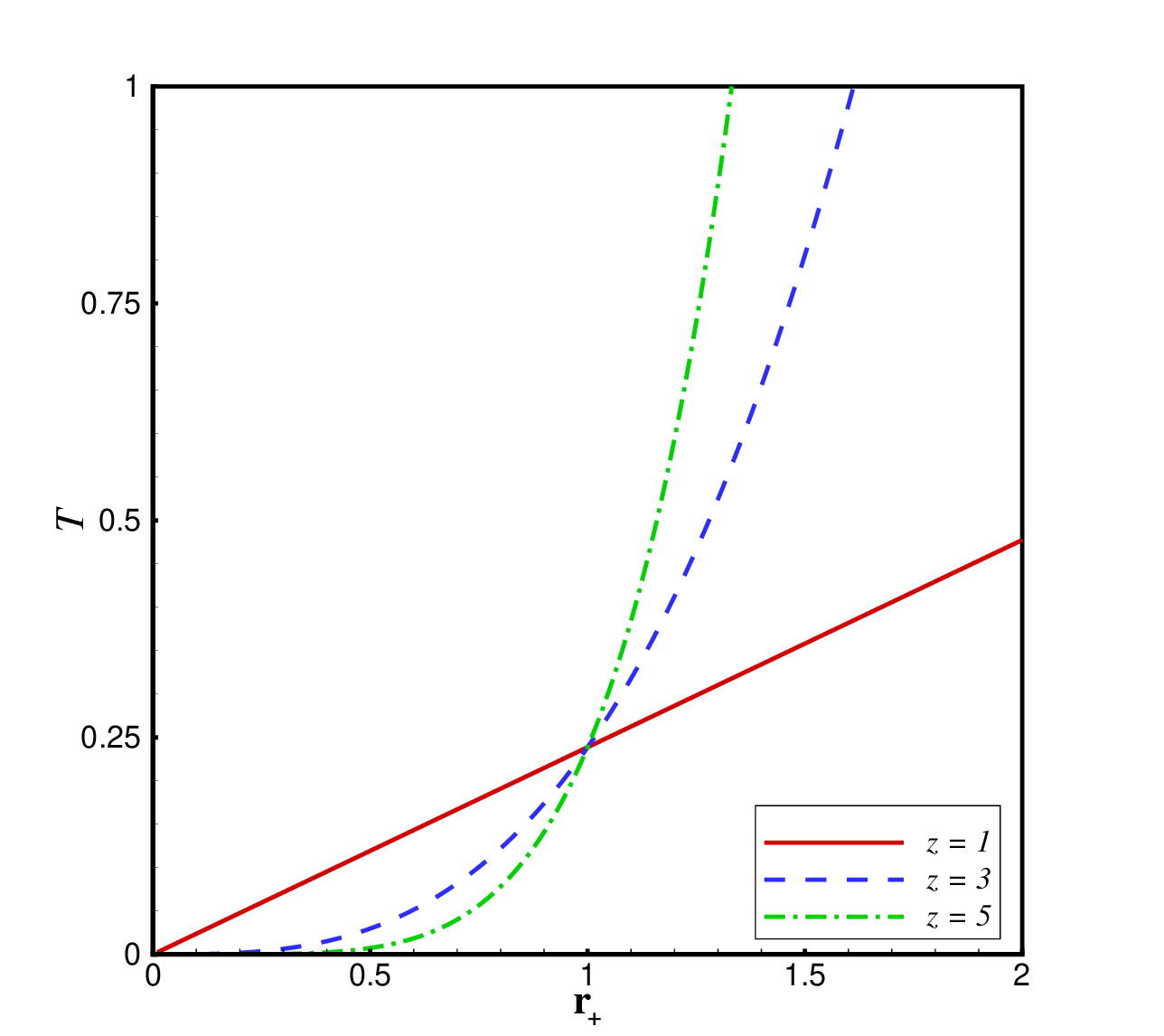}
\caption{\small{Temperature $T$ versus $r_{+}$ with $\eta=3$ and  $l=1$.} \label{fig1}}
\end{figure}

In this section, our purpose is to probe the thermodynamics of the quasitopological Lifshitz dilaton black brane and also obtain the related conserved quantities. At first, we calculate the temperature of this brane by the analytic continuation of the metric. In this method, if we use $t\rightarrow i\tau$, we get to the Euclidean section of the metric \eqref{metric1} which its regularity at $r = r_{+}$ involves us to identify
$\tau\rightarrow \tau+\beta_{+}$, where $\beta_{+}$ is the inverse of Hawking temperature \cite{Bazr3}. By this method, the temperature $T$ at the horizon $r_{+}$ is obtained by
\begin{eqnarray}\label{Temp}
T=\frac{1}{2\pi}\sqrt{-\frac{1}{2}(\nabla_{\mu}\chi_{\nu})(\nabla^{\mu}\chi^{\nu})}\,\mid_{r=r_{+}}=\bigg(\frac{r^{z+1} \sqrt{f^{'}g^{'}}}{4\pi l^{z+1}}\bigg)_{r=r_{+}}=\frac{r_{+}^{z+1}}{4\pi l^{z+1}}\sqrt{f_{1}g_{1}},
\end{eqnarray}
where we have used of the expansions in Eq.\eqref{basthorizon} to obtain the last term. The expansions in Eq.\eqref{basthorizon} show that $f_{1}$ and $g_{1}\propto \frac{1}{r_{+}}$ and therefore the temperature is reduced to
\begin{eqnarray}\label{T1}
T=\frac{\eta}{4\pi l^{z+1}}r_{+}^{z},
\end{eqnarray}
where $\eta$ is a proportionality constant. In order to know more about the temperature behavior, we plot $T$ versus $r_{+}$ for different values of $z$ in Fig. \ref{fig1}. Since physical solutions are only obtained for the positive value of the temperature $T$, so we shlould select a positive value for the constant $\eta$. In Fig.\ref{fig1}, the temperature has a linear behavior versus $r_{+}$ for $z=1$, while it has a curve behavior for $z\neq 1$. So, there is a $r_{+\rm min}$ in which the black brane temperature has the same value for each values of $z$. For $r_{+}<r_{+\rm min}$, the black brane temperature with $z\neq 1$ is smaller than the one with $z=1$ ($T_{z\neq 1}<T_{z=1}$) and for $r_{+}>r_{+\rm min}$, $T_{z\neq 1}>T_{z=1}$.\\ 
In order to earn the entropy of the quasitopological Lifshitz dilaton black brane, we use the Wald formula \cite{Wald} 
\begin{eqnarray}
S=-2\pi \oint d^{n-1} x \sqrt{\tilde{g}} Y^{abcd} \hat{\epsilon}_{ab}\hat{\epsilon}_{cd},
\end{eqnarray}
where $Y^{abcd}=\frac{\partial \mathcal L}{\partial R_{abcd}}$ ($\mathcal{L}$ is the Lagrangian) and $\hat{\epsilon}_{ab}$ is the binormal to the horizon. We integrate on the $(n-1)$-dimensional spacelike hypersurface of the Killing horizon with the induced
metric ${\tilde{g}}_{\alpha\beta}$, whose determinant is $\tilde{g}$. For the quasitopological Lifshitz dilaton black brane, $Y^{abcd}\hat{\epsilon}_{ab}\hat{\epsilon}_{cd}$ is constant on the horizon \cite{Mehdi} and so we can define the entropy density $s=S/V_{n-1}$ where $V_{n-1}$ is the volume of a unit $(n-1)$-sphere. Thus, the entropy density of this black brane can be obtained as bellow \cite{Mehdi}  
\begin{eqnarray}\label{entropy}
s=\frac{r_{+}^{n-1}}{4}.
\end{eqnarray}
In order to calculate the mass of the quasitopological Lifshitz dilaton black brane, we use the counterterm method. For this purpose, we vary of the action $I_{\mathrm{tot}}$ \eqref{Action1} about a solution of the equation of motion  
\begin{eqnarray}
\delta I_{\mathrm{tot}}=\int d^{n}x (S_{ab}\delta h^{ab}+N_{a}\delta U^{a}),
\end{eqnarray}
where $S_{ab}$ and $N_{b}$ are defined by\cite{Deh82,Deh38} 
\begin{eqnarray}\label{Sab}
S_{ab}=\frac{\sqrt{-h}}{16\pi}\{\Pi_{ab}+\frac{2q}{l b^{n-1}}e^{-2\alpha\Phi/(n-1)}(-A_{\gamma}A^{\gamma})^{-1/2}(A_{a}A_{b}-A_{\gamma}A^{\gamma}h_{ab})\},
\end{eqnarray}
\begin{eqnarray}
N_{b}=-\frac{\sqrt{-h}}{16\pi}\{4e^{-4\alpha\Phi/(n-1)}n^{a}F_{a b}-\frac{4q}{l b^{n-1}}e^{-2\alpha\Phi/(n-1)}(-A_{\gamma}A^{\gamma})^{-1/2}A_{b}\},
\end{eqnarray}
and $\Pi_{\alpha\beta}$ is reached as bellow  
\begin{eqnarray}
\Pi_{ab}&=&K_{ab}-Kh_{ab}+\frac{2\lambda}{(n-2)(n-3)}(3J_{ab}-Jh_{ab})+\frac{3\mu}{5n(n-2)(n-1)^2(n-5)}(5H_{ab}-H h _{ab})\nonumber\\
&&+\frac{(n-1)(l^4-2\lambda l^2+3\mu)}{l^5}h_{ab}.
\end{eqnarray}
As the different terms of $N_{b}$ and $S_{ab}$ cancel each other in the background \eqref{lifmet}, so we get to $S_{ab}=0$ and $N_{b}=0$ where they result to $\delta I_{\mathrm{tot}}=0$. Therefor we can conclude that the action $I_{\mathrm{tot}}$ is a finite on-shell one that satisfies a well-defined variational principle for the background metric \eqref{lifmet}.
Now we can compute a finite stress tensor which may carry important physical information for the asymptotically AdS spacetimes that are dual to relativistic field theories \cite{Henni,Balasubramanian}. As the dual field theories for the asymptotically Lifshitz spacetimes are nonrelativistic, so they will not have a covariant relativistic
stress tensor. However, we can specify a stress tensor complex for the asymptotically Lifshitz spacetimes \cite{Ross} that consists of energy density $\mathcal{E}$, energy flux $\mathcal{E}_{i}$, momentum density $\mathcal{P}_{i}$, and
spatial stress tensor $\mathcal{P}_{ij}$ by the bellow formula
\begin{eqnarray}\label{consqua}
\mathcal{E}=2S^{t}{{}_{t}}-N^{t}A_{t},\,\,\,\,\,\,\, \mathcal{E}^{i}=2S^{i}{{}_{t}}-N^{i}A_{t},
\end{eqnarray}
and
\begin{eqnarray}\label{aa}
\mathcal{P}_{i}=-2S^{t}{{}_{i}}+N^{t}A_{i},\,\,\,\,\,\, \mathcal{P}^{j}_{i}=-2S^{j}{{}_{i}}+N^{j}A_{i}.
\end{eqnarray}
The above conserved quantities obey the bellow conservation equations \cite{Deh82,Deh38,Deh44}
\begin{eqnarray} 
\partial_{t} \mathcal{E}+\partial_{i} \mathcal{E}^{i}=0,\,\,\,\,\,\,\,\partial_{t} \mathcal{P}_{j}+\partial_{i} \mathcal{P}^{i}_{j}=0.
\end{eqnarray}
Using Eqs. \eqref{Sab}-\eqref{consqua}, we calculate the energy density of the quasitopological Lifshitz dilaton black brane as this
\begin{eqnarray}\label{Einf}
\mathcal{E}&=& \frac{(n-1)\sqrt{f}(1-\sqrt{g})r^{n+z-1}}{8\pi l^{z+1}}- \frac{\lambda(n-1)\sqrt{f}(1-g^{3/2})r^{n+z-1}}{12\pi l^{z+3}}+\frac{3\mu(n-1)\sqrt{f}(1-g^{5/2})r^{n+z-1}}{40\pi l^{z+5}}\nonumber\\
&&+\frac{q^2 k}{4\pi b^{n-1} l^{z+1}}(e^{-2\alpha\Phi/(n-1)} r^{n-1}-b^{n-1}).
\end{eqnarray}
If we replace the large $r$ expansions \eqref{soluinf} in Eq. \eqref{Einf}, so we can attain the mass of this black brane as the following
\begin{eqnarray}\label{smarr2}
\mathcal{E}=\frac{n-1}{16\pi (n+z-1)}C_{\mathrm{tot}},
\end{eqnarray}
where $C_{\mathrm{tot}}$ is the total constant at the infinity that has been defined in Eq. \eqref{Ctinf}. So, the above relation shows that the mass of the quasitopological Lifshitz dilaton black brane is related to the total constant at the infinity. \\
On the other side, the product of the entropy density $s$ in Eq. \eqref{entropy} and the temperature $T$ in Eq. \eqref{Temp} is related to the total constant at the horizon in Eq. \eqref{Cth} with the bellow relation 
\begin{eqnarray}\label{smarr1}
Ts=\frac{C_{\mathrm{tot}}}{16\pi}. 
\end{eqnarray}
So, we can deduce from Eqs. \eqref{smarr2} and \eqref{smarr1} that 
\begin{eqnarray}
\mathcal{E}=\frac{n-1}{n+z-1}Ts,
\end{eqnarray} 
where by using the relations \eqref{T1} and \eqref{entropy}, the smarr-type formula is derived as 
\begin{eqnarray}
\mathcal{E}=\frac{\eta(n-1)}{16\pi l^{z+1}(n+z-1)}(4s)^{\frac{n+z-1}{n-1}}.
\end{eqnarray}
Now by the evaluated conserved and thermodynamic quantities, we can check the validity of the first law of the thermodynamics. Regard $s$ as an extensive parameter for the mass, $\mathcal{E}(s)$, one can describe the related intensive parameter conjugate to $s$ by
\begin{eqnarray}\label{first}
T=\bigg(\frac{\partial{\mathcal{E}}}{\partial{s}}\bigg).
\end{eqnarray}
It is easy to show that the intensive quantity $T$ in Eq. \eqref{first} coincides with the temperature evaluated in Eq. \eqref{T1}. This shows that the obtained quantities for the quasitopological Lifshitz dilaton black brane satisfy the first
law of the thermodynamics
\begin{eqnarray}
d\mathcal{E}=Tds.
\end{eqnarray} 
\section{Thermal stability}\label{stability} 
\begin{figure}
\center
\includegraphics[scale=0.5]{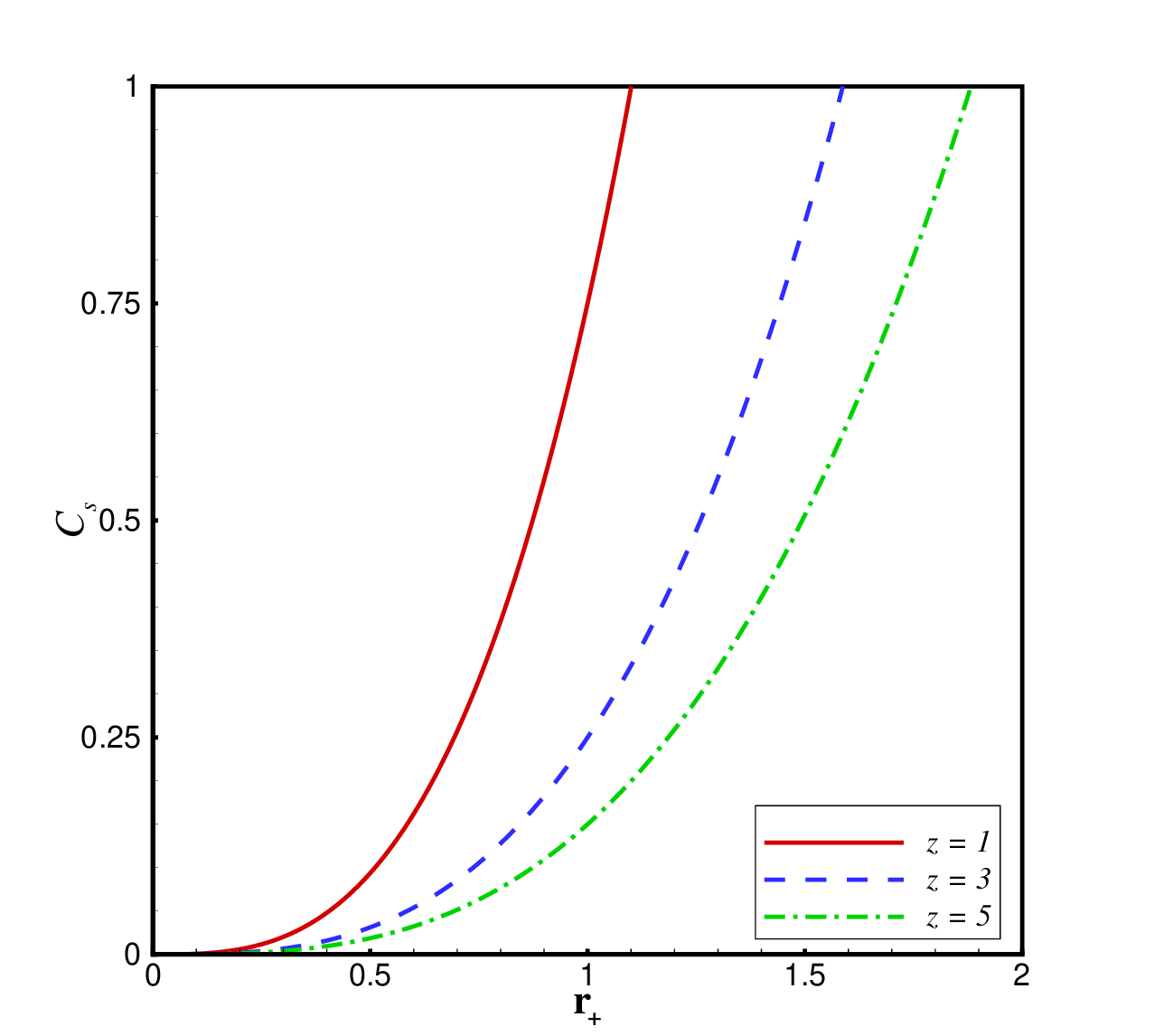}
\caption{\small{Heat capacity $C_{s}$ versus $r_{+}$ with $n=4$ and $l=1$.} \label{fig2}}
\end{figure}

In this section, we are going to study thermal stability of the quasitopological Lifshitz dilaton black brane. One can analyze the stability of a thermodynamic system with respect to small variations of the thermodynamic coordinates by investigating the entropy behavior around the equilibrium. It is also possible to study the thermal stability by using the energy density. So, a thermodynamic system is thermally stable, if the related heat capacity $C_{s}$ is positive. The heat capacity of the uncharged quasitopological Lifshitz dilaton black brane is obtained as bellow
\begin{eqnarray}
C_{s}=T\bigg(\frac{\partial s}{\partial T}\bigg)=T\bigg(\frac{\partial^{2} \mathcal{E}}{\partial s^2}\bigg)^{-1}=\frac{n-1}{4z}r_{+}^{n-1}.
\end{eqnarray}
It is clear that $C_{s}$ is positive for each positive values of $n$ and $z$. In order to manifest this behavior better, we have plotted $C_{s}$ versus $r_{+}$ for different values of $z$ with the smallest dimension $n=4$ in Fig. \ref{fig2}. 
We also remind that a physical solution can be reached, if the condition $T>0$ is established. Therefore, we can conclude that the quasitopological Lifshitz dilaton black brane is thermally stable for each values of $n$ and $z$ with $\eta>0$.
\section{concluding remarks}\label{result}
In this paper, we could obtain a new class of the Lifshitz solutions for a dilaton black brane in the presence of the cubic quasitopological gravity. Quasitopological gravity is one of the higher-order curvature one with the ability to construct second-order differential equations when the metric is spherically symmetric. In AdS/CFT correspondence, quasitopological gravity can provide enough free coupling parameters to make a one-to-one relation
between the central charges on the nongravitational
side and the coupling parameters on the
gravitational side. In spherically symmetric conditions, quasitopological gravity is similar to the LoveLock theory. However, since the quasitopological gravity has some terms which are not true topological invariants, so it can lead to the gravitational effects in more dimensions than the LoveLock theory. For example, as the cubic quasitopological gravity may contribute in the action with $n\geq4$ ($n$ denotes the space dimension), but the third-order LoveLock theory is effective for $n\geq6$.\\
We started our theory with a bulk action including a dilaton field, cubic quasitopological gravity and a massless gauge field and the Lifshitz metic with a $(n-1)$-dimensional
hypersurface that contains a zero curvature boundary. In order to get a finite well-defined action, we generalized the counterterm
method to the quasitopological gravity and added some boundary terms and counterterms with a flat boundary, $\hat{R}_{abcd}=0$. Then, we varied the total action with respect to the functions and obtained the field equations. By imposing the Lifshitz condition, $f(r)=g(r)=1$ on the equations, we attained some constraints on the theory parameters, $\Lambda$, $q$ and $\alpha$ and the dilaton field $\Phi(r)$ with and without matter sources. Under some conditions, it is possible to obtain exact solutions for the field equations, while they are not solved analytically in general. So we used a short cut and found a total constant, $\mathcal{C}_{\mathrm{total}}$ along the radial coordinate $r$ by the equations which could create a relation between the thermodynamic quantities. We also probed the solutions at the horizon and the infinity. At the horizon, the value of the total constant is related to the product of the values of the entropy density $s$ and the temperature $T$. By variation of the total action about a solution of the equation of the motion, we could define a stress tensor complex that included the energy density $\mathcal{E}$. Obtaining the energy density, it is related to the total constant at the large $r$ expansion. So, by considering the total constant as an intermediary, we could provide a smarr-type formula and show the energy density as a function of $s$ and $T$. We also checked the validity of the first law of the thermodynamics and proved that the quasitopological Lifshitz dilaton black brane solutions obey this law. Then we investigated the thermal stability of the quasitopological Lifshitz dilaton black brane. The heat capacity $C_{s}$ is positive for each positive values of the critical exponent $z$ and the space dimension $n$. For $\eta>0$ (where $\eta$ shows a proportionality constant), the temperature is positive. So the cubic quasitopological Lifshitz dilaton black brane is thermally stable. \\
In this paper, we were successful to study the solutions of the static Lifshitz dilaton black brane in the cubic quasitopological gravity. It may be interesting to study the Lifshitz dilaton black brane solutions in the higher-order quasitopological gravity. In our next study, we have a desire to probe the quartic quasitopological dilaton black brane in the Lifshitz spacetime with a rotation. We can also add a massive gauge field to have a larger class of the Lifshitz solutions.

\acknowledgments{{We would like to thank Payame Noor University and Jahrom University.}


\begin{thebibliography}{99}
\bibitem{Ran}L. Randallt and R. Sundrum, "Large mass hierarchy from a small extra dimension." Phys. Rev. Lett. \textbf{83}, 3370 (1999); 	[arXiv:hep-ph/9905221].

\bibitem{Dvali} G. Dvali, G. Gabadadze, and M. Porrati, "4D gravity on a brane in 5D Minkowski space." Phys. Lett. B \textbf{485},
208 (2000).

\bibitem{Kov1} P. Kovtun, D. T. Son, and A. O. Starinets, "Holography and hydrodynamics: Diffusion on stretched horizons."
J. High Energy Phys. \textbf{10} 064 (2003); [arXiv:hep-th/0309213
]. P. Kovtun, D. T. Son, and A. O. Starinets, "Viscosity in strongly interacting quantum field theories from black hole physics." Phys. Rev. Lett \textbf{94}, 111601 (2005); [arXiv:hep-th/0405231].

\bibitem{Buch} A. Buchel, "On universality of stress-energy tensor correlation functions in supergravity." Phys. Lett. B \textbf{609}, 392 (2005); [arXiv:hep-th/0408095].

\bibitem{Lan} K. Landsteiner and J. Mas, "The shear viscosity of the non-commutative plasma." J. High Energy Phys. \textbf{07} 088 (2007); [arXiv:0706.0411 [hep-th]].

\bibitem{Riess} A.G. Riess et al., "Observational evidence from supernovae for an accelerating universe and a cosmological constant." Astron. J. \textbf{116}, 1009 (1998).

\bibitem{Halver} N.W. Halverson et al., "Degree angular scale interferometer first results: a measurement of the cosmic microwave background angular power spectrum." Astrophys. J. \textbf{568}, 38 (2002).

\bibitem{Tonry} J.L. Tonry et al., "Cosmological results from high-z supernovae." Astrophys. J. \textbf{594}, 1 (2003).

\bibitem{Sperg} D.N. Spergel et al."First-year Wilkinson Microwave Anisotropy Probe (WMAP)* observations: determination of cosmological parameters." , Astrophys. J. Suppl. \textbf{148}, 175 (2003).

\bibitem{Sta} A. A. Starobinsky,"A new type of isotropic cosmological models without singularity." Phys. Lett. B \textbf{91}, 99 (1980).

\bibitem{Car} S. M. Carroll, V. Duvvuri, M. Trodden and M. S. Turner, "Is cosmic speed-up due to new gravitational physics?." Phys. Rev. D \textbf{70}, 043528 (2004); [arXiv:astro-ph/0306438].

\bibitem{Fay} S. Fay, R. Tavakol and S. Tsujikawa, Phys. Rev. D \textbf{75}, 063509 (2007).

\bibitem{Amen}  L. Amendola,"Scaling solutions in general nonminimal coupling theories." Phys. Rev. D \textbf{60}, 043501 (1999); [arXiv:astro-ph/9904120].

\bibitem{Brans} C. Brans, R.H. Dicke, "Mach's principle and a relativistic theory of gravitation." Phys. Rev. \textbf{124}, 925 (1961).

\bibitem{Brans2} Dicke R. H. "Mach’s Principle and Invariance under Transformation of Units." Phys. Rev., \textbf{125}, 2163 (1962).

\bibitem{dilaton1} M. B. Green, J. H. Schwarz and E. Witten, "Superstring Theory", Cambridge University Press, Cambridge (1987).

\bibitem{Gregory} R. Gregory and J. A. Harvey, "Black holes with a massive dilaton." Phys. Rev. D \textbf{47}, 2411 (1993).

\bibitem{Gibbons} G.W. Gibbons and K. Maeda,"Black holes in Kaluza-Klein theory." Ann. Phys. (N.Y.) \textbf{167}, 201 (1986).

\bibitem{Horne} J. H. Horne and G. T. Horowitz, "Rotating dilaton black holes." Phys. Rev. D \textbf{46}, 1340 (1992); [arXiv:hep-th/9203083].

\bibitem{Oliva} J. Oliva and S. Ray  "A new cubic theory of gravity in five dimensions: black hole, Birkhoff's theorem and C-function." Class. Quantum Grav. \textbf{27}, 225002 (2010); [arXiv:1003.4773 [gr-qc]].

\bibitem{Mye} R.C. Myers and B. Robinson, "Black holes in quasi-topological gravity." J. High Energy Phys. \textbf{08}, 067 (2010); [arXiv:1003.5357 [gr-qc]].

\bibitem{Love1} D. Lovelock, "The Einstein Tensor and Its Generalizations." J. Math. \textbf{12} 498 (1971).

\bibitem{Buchel} A. Buchel and R.C. Myers, "Causality of holographic hydrodynamics." J. High Energy Phys. \textbf{08},
016 (2009); [arXiv:0906.2922 [hep-th]]. M. Brigante, H. Liu, R. C. Myers, S. Shenker and S. Yaida,"Viscosity bound and causality violation." Phys. Rev. Lett. \textbf{100}, 191601 (2008); [arXiv:0802.3318 [hep-th]]. X.O. Camanho and J.D. Edelstein, "Causality constraints in AdS/CFT from conformal collider physics and Gauss-Bonnet gravity." J. High Energy Phys. \textbf{04}, 007 (2010); [arXiv:0911.3160].

\bibitem{Escobedo} X.H. Ge and S.J. Sin, "Shear viscosity, instability and the upper bound of the Gauss-Bonnet coupling constant." J. High Energy Phys. \textbf{05}, 051 (2009); [arXiv:0903.2527 [hep-th]]. R.G. Cai, Z.Y. Nie
and Y.W. Sun, "Shear viscosity from effective couplings of gravitons." Phys. Rev. D 78, 126007 (2008); [arXiv:0811.1665 [hep-th]]. A. Buchel, J. Escobedo, R.C. Myers, M.F. Paulos, A. Sinha and M. Smolkin, "Holographic GB gravity in arbitrary dimensions." J. High Energy Phys. \textbf{03}, 111 (2010).

\bibitem{Yang} X.O. Camanho and J. D.
Edelstein, "Causality in AdS/CFT and Lovelock theory." J. High Energy Phys. 06, 099 (2010);X.H. Ge, S.J. Sin, S.F. Wu and G.H. Yang,"Shear viscosity and instability from third order Lovelock gravity." Phys. Rev. D \textbf{80}, 104019 (2009) [arXiv:0905.2675 [hep-th]];

\bibitem{Myers2} R.C. Myers, M.F. Paulos, A. Sinha,"Holographic studies of quasi-topological gravity." J. High Energy Phys. \textbf{08}, 035 (2010) [arXiv:1004.2055 [hep-th]].

\bibitem{Hofman} D. M. Hofman, "Higher derivative gravity, causality and positivity of energy in a UV complete QFT." Nucl. Phys. B \textbf{823}, 174 (2009); [arXiv:0907.1625 [hep-th]].

\bibitem{Kuang} X. M. Kuang, W. J. Li, Y. Ling, "Holographic superconductors in quasi-topological gravity." J. High Energy Phys. \textbf{12} 069 (2010); J. P. Wu,  "Holographic fermions in charged Gauss-Bonnet black hole." J. High Energy Phys. \textbf{07} 106 (2011); X. M. Kuang, W. J. Li, Y. Ling,"Holographic p-wave superconductors in quasi-topological gravity." Class. Quantum. Grav \textbf{29} 085015 (2012); [arXiv:1106.0784 [hep-th]].

\bibitem{Brenna1} W. G. Brenna, M. H. Dehghani, and R. B. Mann,"Quasitopological Lifshitz black holes." Phys. Rev. D \textbf{84}, 024012 (2011); [arXiv:1101.3476 [hep-th]].

\bibitem{Brenna} W. G. Brenna and R. B. Mann, Quasitopological Reissner-Nordström black holes, Phys. Rev. D \textbf{86} 064035 (2012); [arXiv:1206.4738].

\bibitem{Mehdi} M. H. Dehghani, A. Bazrafshan, R. B. Mann, M.
R. Mehdizadeh, M. Ghanaatian and M. H. Vahidinia, Black Holes in (Quartic) Quasitopological Gravity, Phys. Rev. D \textbf{85} 104009 (2012); [arXiv:1109.4708].

\bibitem{Bazr3} M. Ghanaatian, F. Naeimipour, A. Bazrafshan and M. Abkar, Charged black holes in quartic quasi-topological gravity, Phys. Rev. D \textbf{97} 104054 (2018); [arXiv:1801.05692 [gr-qc]].

\bibitem{ahmadi} M. Ghanaatian, F. Naeimipour, A. Bazrafshan, M. Eftekharian, and A. Ahmadi, Third order quasitopological black hole with a power-law Maxwell nonlinear source. Physical Review D, \textbf{99}, 024006 (2019); [arXiv:1809.05198 [gr-qc]].

\bibitem{Ghan1} M. Ghanaatian, A., Bazrafshan, and W. G. Brenna, Lifshitz quartic quasitopological black holes, Physical Review D, \textbf{89} 124012 (2014).

\bibitem{kord1}M. Kord Zangeneh, M.H. Dehghani, A. Sheykhi, Thermodynamics of Gauss–Bonnet-dilaton Lifshitz black branes, Phys. Rev. D \textbf{92} (2015) 064023; [arXiv:1506.07068 [hep-th]]

\bibitem{Kachru} S. Kachru, X. Liu, and M. Mulligan, Gravity duals of Lifshitz-like xed points, Phys. Rev. D \textbf{78}, 106005
(2008); [arXiv:0808.1725].

\bibitem{Tarrio}J. Tarrio, S. Vandoren, Black holes and black branes in Lifshitz spacetimes, J.High Energy Phys. \textbf{09} 017 (2011); [arXiv:1105.6335 [hep-th]].

\bibitem{Bert} G. Bertoldi, B.A. Burrington, A.W. Peet, Thermal behavior of charged dilatonic black branes in AdS and UV completions of Lifshitz-like geometries, Phys. Rev. D \textbf{82} 106013 (2010); arXiv:1007.1464 [hep-th].

\bibitem{Deh82} M. H. Dehghani and R. B. Mann, Thermodynamics of Lovelock-Lifshitz Black Branes, Phys. Rev. D \textbf{82} 064019 (2010); 
[arXiv:1006.3510 [hep-th]].

\bibitem{Deh38} M. H. Dehghani and Sh. Asnafi, Thermodynamics of rotating Lovelock-Lifshitz black branes, Phys. Rev. D \textbf{84} 064038 (2011); 
[arXiv:1107.3354 [hep-th]].

\bibitem{Deh44} M. H. Dehghani and M. H. Vahidinia, Surface terms of quasitopological gravity and thermodynamics of charged rotating black branes, Phys. Rev. D \textbf{84} 084044 (2011) [arXiv:1108.4235 [hep-th]].

\bibitem{Ross} S. F. Ross and O. Saremi,"Holographic stress tensor for non-relativistic theories", J. High Energy Phys. \textbf{09} 009 (2009) [arXiv:0907.1846 [hep-th]].


\bibitem{Wald} R. M. Wald, "Black hole entropy is the Noether charge," Phys. Rev. D \textbf{48}, R3427 (1993) [arXiv:gr-qc/9307038].

\bibitem{Henni} M. Henningson and K. Skenderis, "The holographic Weyl anomaly." J. High Energy Phys. \textbf{07} 023 (1998) [arXiv:hep-th/9806087].

\bibitem{Balasubramanian} V. Balasubramanian and P. Kraus,"A stress tensor for anti-de Sitter gravity," Commun. Math. Phys. \textbf{208}, 413 (1999), [	arXiv:hep-th/9902121].

\end{thebibliography}
\end{document}